# Can elemental bismuth be a liquid crystal?

Nathan Argaman[*]

Physics department, N.R.C.N., P.O. Box 9001 Be'er Sheva 84190, ISRAEL.

(March 28, 2010)

A number of anomalies have been reported in molten Bi, including a first-order liquid-liquid transition at 1010K and ambient pressure, which is irreversible at cooling rates of several degrees per minute. An interpretation of these effects as due to long-range orientational order is suggested. Significant evidence for directionality in liquid Bi, albeit only immediately after melting, is available in experiments made circa 1930. Further experimentation is called for.

---

[*] E-mail: argaman@mailaps.org .

It is well known that many pure elements exhibit thermodynamic first-order phase transitions in the solid state, abruptly transforming from one crystalline arrangement to another in response to changes in temperature and/or pressure. During much of the 20$^{th}$ century, it was generally thought[1,2] that once an element is heated above melting, the only first-order transition it exhibits is the liquid-gas transition. However, in the 90's, Brazhkin and coworkers[3] found high-pressure first-order transitions in melts of several elements (Se, S, Bi, Sn, Sb, Te, and I), with, in some cases, several transitions observed in the same element. First-order transitions in additional elements (e.g., in P and N) have been identified and studied subsequently.[4,5] Such transitions are invariably attributed to changes in the short-range order of the liquid.[3] As discussed below, this scenario is appropriate for transitions which exhibit a negative slope in the corresponding pressure–temperature phase diagram, and indeed most of the observed transitions do. The purpose of the present work is to suggest an alternative scenario, involving long-range orientational order, for some of the remaining transitions. In other words, the possibility that liquid-crystal regions occur in the phase diagrams of the elements will be considered, focusing on the case of Bi.

One might initially object to the notion of liquid-crystal behavior in an elemental melt, because the individual atoms are isotropic entities, and do not have the needle-like or plate-like shapes generally associated with mesogens (liquid-crystal forming molecules). However, in many cases the ordering of a liquid crystal is due primarily to the bonds between the constituents, rather than the geometry of the constituents themselves, as in amphiphilic substances forming smectic phases. In covalently bonded elements, one would indeed generally expect the bonds to dominate.

The motivation for focusing on Bi is due to two recent ambient-pressure observations: (a) Greenberg et al.[6] observed clear signatures of a first-order phase transition in molten Bismuth, including a discontinuous drop in density of about 1%, at a temperature of 1010K (this is quite remote from both the melting point, 544K, and the boiling point, 1837K). These measurements appear to reveal additional, weaker anomalies at other temperatures. (b) Li et al.[7] used resistivity measurements to obtain evidence for transformations in melts of Bi, Sb and Sn and some of their alloys at temperatures somewhat higher than 1000K. At the scanning rates of several degrees per minute used in these experiments, the transformations in Bi and Sb were observed to be irreversible, whereas that in Sn was reversible, with hysteresis.

We begin the discussion by considering an element which has a tendency to form two distinct isotropic liquid phases – differing in their short-range structure – which are immiscible, i.e., allow for a first-order transition between them. The phase with a lower internal energy will be stable under normal conditions. If the higher-energy phase is also denser, it can be stabilized by applying pressure. At the transition pressure, latent heat, $T\Delta S$ per mole, must be supplied, to account for the difference in internal energy. The Clausius-Clapyron relationship, $dP/dT = \Delta S/\Delta V$, then dictates that the slope of the transition must be negative, as is the change in molar volume $\Delta V$. This implies that if the transition is crossed by changing temperature rather than pressure, one obtains contraction upon heating, as in melting ice. Such behavior is indeed observed[3,8] in many of the elements in which transitions have been found, including Se, S, Sn, Te and P.

Perhaps the best-studied example is that of phosphorus,[4,8] which at low pressures forms tetrahedral $P_4$ molecules. These nano-pyramids are the building blocks of the solid white phosphorus phase and its melt. They occupy a relatively large volume in the melt, and at a pressure near 1GPa they "collapse," i.e., flatten into kite-shaped $P_4$ units, which connect to form a high-density, networked, metallic fluid, similar to the glassy red-phosphorus phase observed at intermediate pressures at lower temperatures. At $1000^O C$, the density change at the liquid-liquid transition is by a factor of 1.6.

The Clausius-Clapyron argument described above is generic. It covers transitions which are accompanied by metallization[3] or by polymerization,[4] as indeed occurs in some of the examples mentioned, including that of phosphorus. It is not a compelling argument, though, as demonstrated by the liquid-gas transitions, which have positive slopes although they are associated with changes in short-range order (accordingly, they may be driven by increasing temperature or *decreasing* pressure). However, in these transitions the typical changes in density are even more dramatic. Thus, for a liquid-liquid transition with a positive slope and only a 1% change in density, it seems appropriate to consider alternatives to the "change in short-range order" scenario.

When long-range order is present, different phases belong to discrete symmetry classes. As each pressure–temperature point is associated with one such class or another, transitions between them must occur along sharp lines in the pressure–temperature diagram, even if the changes in volume and entropy are small. Such transitions may in principle be either first-order or continuous, but they can not simply end at a critical point, like the liquid-gas transition. Phases with long-range positional order would be solid, but phases with long-range orientational order, i.e., liquid crystals, are not ruled out by the observation of melting. In fact, in many liquid-crystal forming substances, the strong ordering of the solid is lost in a gradual step-by-step manner upon heating, with several phases of decreasing order – both long-range orientational order and short-range positional order – separating between the solid and the isotropic liquid phase.[9]

Consider then the crystal structure of solid Bi, and melting of this structure. Bi has the A7 trigonal structure (as in As and Sb), which is best viewed as a distorted simple cubic lattice, where the (111) planes have been paired, with each pair of adjacent planes forming a bilayer, and with the distances between the bilayers increased significantly.[10] The geometry within the bilayers is roughly as in the undistorted lattice: each atom has three nearest neighbors at a separation of ~3.0Å, with covalent bonds at approximately 90$^O$ to each other, and an interlayer separation of ~1.5Å. The distance between the bilayers is ~2.5Å, giving a next-nearest-neighbor distance of ~3.5Å, with much weaker bonding (as Bi is a semimetal with ~$10^{-5}$ free carriers per atom at low temperatures, very weak metallic bonding may be expected in addition to van der Waals bonding).

It is natural to expect the much-weaker bonds to melt at a relatively low temperature, significantly lower than that which would completely disrupt the ordering imposed by the covalent bonds. Indeed, judging solely from the structure of the solid, one would expect the Bi crystal to loose its symmetry in three discrete steps: At some temperature $T_1$, each bilayer would become sufficiently mobile with respect to the adjacent bilayers to loose registry with them, leading to a hexatic liquid crystal phase; One would expect the bilayers to possibly loose their orientational correlations with each other only at a higher temperature $T_2$, which would signal the transition into a smectic phase. Melting of the smectic phase would occur at a significantly higher temperature, $T_3$. Further complications, including kinetic considerations, would arise due to the presence of the boundaries between the liquid-crystal "grains" or domains. Such effects could perhaps explain the weaker anomalies observed at intermediate temperatures in Ref. 6.

It is thus seen that the tendency of the heavier nitrogen-group elements to form three covalent bonds which are approximately orthogonal to each other, corresponding to the $p_x$, $p_y$ and $p_z$ orbitals, leads to a very asymmetric situation – the region around each atom is divided, roughly speaking, into a "bonding hemisphere" and a "neutral hemisphere." This is more than is needed for the formation of a liquid crystal. Indeed, the layering in Bi is similar to that of the abovementioned amphiphilic molecules, which tend to self-assemble into bilayers, forming smectic liquid crystals. Melting of such a smectic liquid crystal into an isotropic liquid phase is expected to be a first-order transition, in contrast to other liquid-crystal transitions, e.g., the nematic-to-isotropic transition, which are generically continuous.

It is significant to note that the formation of a liquid-crystal phase consisting of stacked bilayers upon cooling from an isotropic liquid phase can easily be frustrated. The ordering within each layer is hexagonal, and as is well known from studies of fullerenes, incorporation of defects – pentagons and heptagons – into such a sheet will be accompanied by folding. If such a sheet folds into a roughly spherical or cylindrical geometry during its formation, it will have reached a metastable state which is likely to be extremely long lived. On the other hand, it is possible that the interatomic forces within the growing bilayer will be such that such topological defects will be efficiently repelled from the bulk of the layer, towards its growing boundary. Furthermore, the presence of a screw dislocation with a Burgers vector perpendicular to the bilayers may significantly increase the growth rate of the liquid crystal region, as it does for solids. In this case, the bilayers are formed in an already-stacked arrangement (perhaps a better description would be in terms of a single bilayer which overlays itself repeatedly). The transformation from a liquid crystal into an isotropic liquid phase upon heating should thus be expected to be largely irreversible in some materials – those in which the bilayers do not have a strong tendency to remain planar – but possibly reversible in others.

Further evidence in favor of the liquid-crystal interpretation may exist in studies of alloys. After all, if Bi forms a liquid crystal over a wide range of temperatures, this type of ordering would not be immediately destroyed by admixture of other elements. In a study[11] of an 80%Bi - 20%Pb alloy, an irreversible transformation was observed above 800K, with a particularly striking signature in the thermopower data (see their Fig. 2). In addition, resolidification of the alloy gives quite different results depending on whether the melt has been overheated to temperatures high enough to undergo the irreversible transformation (see their Fig. 5). In the micrograph of the alloy solidified from the melt which had remained in the temperature range for which the liquid-crystal interpretation is suggested here, apparent correlations of orientations over distances of several millimeters may be tentatively identified. No such correlations are exhibited in the case in which the melt had been heated sufficiently to irreversibly transform into the isotropic liquid state. To substantiate this identification, one would need to examine a statistically significant sample of micrographs.

The proposed interpretation of the phase transition at 1010K in Bi as a smectic-to-isotropic transition would make it easy to explain many of its features: the positive slope, the small change in density, irreversibility, and the presence of additional features at lower temperatures. Of course, various caveats can be raised against it. Much of the experimental evidence has not been corroborated.[12] Irreversible transformations upon heating well above the melting point have been observed in other materials, including alloys such as AlSi, in which no layered structure is expected.[13] These observations have been interpreted in terms of small long-lived solid particles of different compositions, which may exist in the melt as a micro-emulsion, and which only finally "dissolve" or melt at a higher temperature. Although this interpretation is hardly on a firm footing, it is in principle possible that the results for the abovementioned BiPb alloy could be explained in terms of such a mechanism (or a different one), but it is difficult to see how this type of explanation can be extended to pure elements. Other caveats can be raised, e.g., an interpretation of the structure of the Bi melt which is closer to the high-pressure, non-layered Bi-II phase, as seems to be called for from the high density of the melt. In fact, much information regarding the structure of the Bi melt has been obtained from diffraction experiments (see, e.g., Ref. 6) and awaits a detailed interpretation. However, if the liquid-crystal interpretation is to be examined, it would be highly desirable to redesign these experiments so as to obtain diffraction data from mono-domain samples.

At this point, it would appear that all of the above arguments are merely suggestive, and that much additional work is necessary in order to substantiate or disprove the proposed interpretation. Application of theories developed for liquid crystals is called for, and MD and/or DFT simulations could be used to shed light on melting in Bi.[14] A decisive proof, however, would need to be experimental, and may be pursued by adapting the tools used in liquid crystal research[15] to pure elements such as Bi, and to the relatively high temperatures involved.

Before embarking on new experiments, it is appropriate to scan the literature. In fact, striking evidence for orientational order in molten Bi appears in experimental observations made about 80 years ago, in a small temperature interval above melting. At the time, Goetz performed a study[16] of the influence of different factors on the orientation of Bismuth crystals grown from the melt, including "seeding," i.e., placing a small crystal of known orientation in contact with the cooling melt. In many cases, the seed contained twin regions, and it was found that the crystal grew in the twin orientation even if the twin region was completely melted. To verify that molten Bi is capable of remembering its crystal orientation, Goetz heated a polycrystalline rod to a temperature slightly above melting, and observed that upon solidification, "a polycrystalline rod was obtained which had the same position and the same orientation of the different prominent crystal elements as it had before it entered the furnace although it had been molten and recrystallized." Furthermore, the influence of a magnetic field (20,000 Gauss) on the orientation of crystals grown in the absence of a seed was significant, and was interpreted as indicating that the field could "orient" the melt before solidification. The author went as far as using the term "liquid crystal" (with quotation marks in the original) to describe his observations, but eventually preferred an interpretation in terms of a "block phase," in which submicron-sized grains with a crystalline arrangement of the atoms were taken to "exist within a state of equilibrium of dissociation with the liquid metal at each temperature," without long-range order. However, such a mechanism would be even more far-fetched than accepting the existence of a thermodynamically stable liquid-crystal phase in Bi, as surface tension would be expected to drive down the volume of the suggested submicron grains, and it would be difficult to understand how they could remain in dynamic equilibrium with the melt. Furthermore, the orientations of grains of diameter $d$ (and density $\rho$) in a melt of viscosity $v$ would have random angular thermal velocities $\dot{\theta} \sim \sqrt{k_B T / \rho d^5}$ with a relaxation time $\tau \sim \rho d^2 / v$, thus undergoing Brownian motion with an angular diffusion coefficient $D_\theta \approx \tau \dot{\theta}^2 \sim k_B T / v d^3$. Sub-micron grains would loose the "memory" of their orientations in less than seconds.

The experimental work of Goetz described above is by no means isolated. Earlier[17] and later[18] work by other researchers showed that the orientation of melted Bi single-crystals affected the measured thermoelectric power of the melt up to 8 degrees above the melting point. Additionally, the possibility of melting and resolidifying Bismuth without loss of memory of the orientation was corroborated by others.[19] Nevertheless, it seems that these experimental findings were hardly, if at all, referred to after the second world war.

It is important to emphasize that the experimental evidence described, which is limited to a few degrees above melting, concerns memory of the orientation the Bi crystals had before melting, and vice versa: the possibility of magnetically influencing the melt to adopt an orientation which would be "memorized" and remain in place through the solidification transition. This "memory-state" appears to be metastable, as evidenced by the effect of small mechanical disturbances, which can "erase" the memory.[19] A reasonable interpretation of this is in terms of the ease with which the preferred orientations of liquid crystals can be rotated. In fact, it is surprising that the orientation of the solid crystal can survive the melting transformation in Bi, which is accompanied by a reduction of some 10% in volume (though it is possible that not all of this reduction occurs immediately upon melting). The dynamics of the orientation within liquid crystal phases is typically very sensitive to various forces – electric, magnetic, elastic... – and in the case of melting of a polycrystal could also exhibit effects due to domain boundaries (the molten grain boundaries). It is therefore natural to expect a liquid-crystal phase to be thermodynamically stable also at temperatures higher than those in which the "memory" effect was observed.

The findings of Goetz and his contemporaries appear to be related to a striking effect reported in modern calorimetric studies of resolidification of Bismuth after heating slightly above melting.[20,21] These studies have not noted the possibility of such a relation, and have accordingly made no attempt to monitor crystal orientations. They have identified a transition between "equilibrium" and "explosive" resolidification, with negligible supercooling of the melt in the first case, and significant supercooling in the second. It appears safe to assume that memory of the orientations can not be maintained in the "explosive" case, in which release of latent heat causes the cooling sample to very rapidly re-heat, returning to the melting temperature. Thus, the orientational effects observed by Goetz can be expected only in samples exhibiting "equilibrium" resolidification. It should be noted that the studies quoted make conflicting claims regarding the reported transition: Aleksandrov and coworkers[20] report a sharp "phase transition" between "equilibrium" and "explosive" behaviors, regardless of heating and cooling rates (although the temperature associated with this transition differs between the earlier and the later publications); Tong and Shi[21] report a transient effect, with only "explosive" resolidification observed if the sample is held at temperatures above melting for an extended time. Clearly, some of the relevant parameters are not under sufficient experimental control to achieve reproducibility of the phenomena.

In terms of a liquid-crystal interpretation, one is prompted to identify the phase which displays "equilibrium" resolidification as a metastable, hexatic phase, with the "explosively" resolidifying melt (up to 1010K) identified as smectic. This identification must be viewed as very tentative, pending an understanding of the additional anomalies observed in Bi at intermediate temperatures – not only in Ref. 6, but also, e.g., in the temperature-dependence of the sound velocity[22] and internal friction[23] – and of the high-pressure measurements of Bi of the early 1990's, in which three distinct liquid phases were identified.[24,3]

In summary, a line of inquiry involving long-range orientational order in elemental melts has been suggested. The attractiveness of this suggestion is due to a comparison of the multitude of phase transitions and reorientation effects associated with liquid crystals, on one hand, with the variety of anomalies observed in molten Bi, which would otherwise remain mysterious, on the other. Particular attention has been given to the liquid-liquid phase transition at 1010K, explaining its positive slope, relatively small density jump, and its apparent irreversibility, and to the orientation-memory effects near melting, i.e., just above 544K. It has been argued that progress in this direction would have to be experimental, and could be achieved, e.g., by seeking directionality in molten Bi using polarized light. It is necessary to find out whether the phenomena observed in the past[16,20] are indeed due to the formation of a liquid crystal phase in Bi, and to identify the specific phase or phases (hexatic? smectic?) and range of stability. Merely reproducing these phenomena in an experimentally well-controlled fashion would constitute a significant advance, even if the resulting understanding would characterize them as metastable or transient effects. Finding a liquid-crystal phase to be thermodynamically stable in an elementary melt, even over only a very limited range in temperature and pressure, would be a revolutionary change in our understanding of the phase diagrams of the elements. If the range of stability of liquid-crystal phases in simple melts is much wider (Sb and Sn are additional candidates[25,26,3,7,23]), then this could serve as a basis for solving many of the riddles in the field of liquid-liquid transformations, and would likely have significant technological implications as well.[11]

The author wishes to thank Y. Greenberg, E. Yahel and G. Makov for many intriguing conversations, and for sharing unpublished data. Use of the PROLA database is also gratefully acknowledged – in including research from early years in this widely available tool, the APS has made an important step towards disseminating the knowledge of physics.